\magnification=1200
\def\square{\mathop{\mkern0.5\thinmuskip
                 \vbox{\hrule
			        \hbox{\vrule
                             \hskip5pt
                             \vrule height5pt width 0pt
                             \vrule}%
                       \hrule}%
                 \mkern0.5\thinmuskip}}
\def\=#1{{\bar {{\bar #1}\!\ }}}
\def\pol{{1\over 2}}
\def\S{S^{(2)}}
\def\mi#1#2{\mu ({#1}_{#2})}
\def\mii#1#2#3{\mi {#1}{#2#3}}
\def\\{\backslash}
\def\x#1{#1\in\S}
\voffset1.8truecm
\hoffset1.9truecm
\hsize 12.5true cm
\vsize 20 truecm
\baselineskip0.4truecm
\def\Marekbox{\vtop to0pt{\bigskip\vskip0.5cm
              \hbox{\it Czachor}\vss}}
\footline={\hfil%
   \hbox to0pt{%
      \hbox to2truecm{\hss%
                  \Marekbox%
                  \llap{\tenrm\folio}
                 }\hss}}

\headline={\hfil}
\hrule height0pt
\vskip1.3truecm
\font\3=cmsl8
\leftline {\bf ON CLASSICAL MODELS OF SPIN}
\vskip1.726truecm
{\parindent=2.1truecm
{\obeylines
 {\bf Marek Czachor}\vskip0.424truecm
 {\3 Centre for Theoretical Physics

 Polish Academy of Sciences

Al. Lotnik\'ow 32/46, 02-668 Warszawa, Poland\/}\ $^1$
\vskip0.848truecm
 Received September 17, 1991; revised  March 17, 1992
}}
\vskip0.848truecm
\noindent
  We discuss two classical situations that   lead   to   probabilities
characteristic for systems with spin-1/2.
(a) Pitowsky model:
It is   demonstrated   that   the   definition   of
spin functions does not imply which circle (a parallel or a great
circle) on the sphere should be taken as a probability   space   in
calculation of conditional probabilities.   Pitowsky's   choice   of
parallels must be formulated as an assumption about the model. It
is shown that the model explicitly avoiding   this   difficulty   is
possible and no contradiction with the Bell Theorem is found. The
modification is based on a new pathological decomposition of   the
sphere and belongs to a class of hidden   variable   theories   with
undetected signals.
(b)  Aerts model:
 We   show   the   importance   of   the   ``polarization
effect" of the measurements  for the sake of   obtaining
a non-Kolmogorovian probability model. It is also shown that   the
conditioning by a   change   of   state   leads   in   general   to   the
non-Kolmogorovian probability calculus.
\bigskip
\item{\bf 1.} {\bf INTRODUCTION}
\bigskip
  This paper is devoted   to   an   analysis   of   two   hidden variables
models   of   spin-1/2 [1,2].   Both   of   them   are   based   on     an
observation   that   a   structure   of     conditional     probabilities
characteristic for systems with spin is not a Kolmogorovian   one.
The problem   is   rooted   in   a   non-Bayesian   structure   of   such
probabilities, and is typically manifested by a violation of   the
Bell inequality. Therefore any model   of   spin   must   violate   at
least one assumption necessary   for   a   derivation   of   the   Bell
inequality.

The purpose of this paper is manifold. First, I will   carefully
re-examine the probability structures of   the   models   and   focus
on some aspects that have not been   examined   in   detail   by   the
authors. A detailed analysis of the Pitowsky   model   will   reveal
several delicate   problems   of   Gudder's   theory   of   probability
manifolds [3]. Variations on the Aerts model will show   that   its
non-Kolmogorovity is explicit, and the change   of   state   of   the
system by the measurement is a necessary aspect of the   model   in
relation     to     the     non-Kolmogorovian       character of       its
probabilistic description.   The   fifth   variation   serves   as   an
example of a macroscopic system whose   non-Kolmogorovity   follows
only from the conditioning by a change of state and there   is   no
lack of knowledge about the measurement.
\bigskip
\item{\bf 2.} {\bf THE PITOWSKY MODEL}
\bigskip
  Let $\S$ be the   unit   sphere.   Pitowsky   proved   [1]   that   it   is
possible to cover $\S$      with ``white"   and   ``black"   points   (=   to
define a function  $S: \S\to \{-1,1\}; x\in \S$  is white   if    $S(x)=1$)
in such a way that

(a) opposite poles have different colours ($S(-x)=- S(x)$),

(b) a probability measure of the set of
white   points   on   a   parallel   whose
lattitude measured from a white pole is  $\theta$,
 is equal to $\cos^2(\theta/2)$.

Strictly speaking the above concise version of the   model   is
not identical to that given in [1],   but   this,   equivalent   form
shall be convenient for our purposes.

In what follows I shall prove some results   essential   for   the
probabilistic interpretation of the ``Pitowsky sphere."

Let $S_+$   denote a set of white points on the Pitowsky sphere  $S$
i.e., $S_+  =\{x\in  \S |  S(x)=1\}$, and let $H_x\subset \S$
 be a hemisphere with   a
pole $x$.
\bigskip

 {\it Lemma 1.\/}
  $\exists_{x\in\S}$ $S_+\cap H_x$ is
    non measurable   in   terms   of   a   Lebesgue
  measure on $\S$.
  \bigskip
  {\it Proof}.
       Assume  {\it ad absurdum\/}   that
	    $$\forall_{\x x}  S_+\cap H_x$$
	     is $\mu$-measurable.   Let
$H_{x+}=S_+\cap H_x$. Its measurability is assumed to hold for all   $x$,
    so
restricting in no way the generality of our considerations we can
take some white $x$. Hence
$$
\mii Hx+ = 2\pi\int_0^{\pi/2}\sin\theta\cos^2(\theta/2)d\theta
={3\over 2}\pi.\eqno(1)
$$
  Consider now any black point $\x y$. Analogously
  $$
  \mii Hy+ =\pol \pi.\eqno(2)
  $$
Let   us   define   now   the   following   sets:
$$A_x=H_x\\H_y,\ A_y=H_y\\H_x,\ A_{x+}=A_x\cap S_+,\
A_{y+}=A_y\cap S_+.
$$
 Let an angular   distance    $\delta(x,y)= \varepsilon$     between
the points $x$ and $y$ be $0<\varepsilon  < \pi$. $A_x$   and $A_y$
   are measurable and
   $$
   \mi Ax=\mi Ay=2\varepsilon.\eqno(3)
   $$
For any two  $\mu$-measurable sets $A$ and $B$, one has
$\mu(A)=\mu(A\\B)+\mu(A\cap B)$
hence
$$
\mu(A) - \mu(B)=\mu(A\\B) - \mu(B\\A).\eqno(4)
$$
We therefore have  $\mii Hx+ - \mii Hy+=\mii Ax+ - \mii Ay-$.
 From   (3)   we   get
 $\mii Ax+\le 2\varepsilon,\ \mii Ay+\le 2\varepsilon$, so
 $$
 |\mii Hx+ - \mii Hy+|\le 2\varepsilon.\eqno(5)
 $$

Inserting (1) and (2) into (5), we obtain  $\varepsilon\ge\pi /2$.

On the Pitowsky sphere a probability   measure   of   a   set   of
black points on a parallel of lattitude  $\varepsilon$
   around a white pole   is
given by $\sin^2(\varepsilon/2)$.
 Therefore in any   neighbourhood   of   a   white
pole   there   exists   a   black   point,    hence
$\varepsilon\ne 0$
   can   be   taken
arbitrarily   small   which   is   in   contradiction   with   the   last
inequality.$\square$
\bigskip
  A straightforward consequence of Lemma 1  is the following
  \bigskip
       {\it Theorem 1.\/}
  $S_+$   is nonmeasurable in terms of the Lebesgue measure on $\S$.
\medskip
{\it Proof}.
       Let $S_+$   be measurable.   According   to   Lemma 1
	        there exists
  $\x x$       such that   $S_+\cap H_x$     is   nonmeasurable.
     Therefore   $H_x$     is
nonmeasurable.     However    $H_x$
       is     measurable     for     any     $x$.
Contradiction. $\square$
\medskip
         Let now   $\bar p$
	       denote   an   outer   probability   measure   on   $\S$
(normalized       outer       Lebesgue       measure).       Moreover
         let
$K_{x,\theta}=\{\x y | \delta(x,y)$ $\le \theta,\ S(x)=+1\}$, $
c_{x,\varphi}=\{y\in K_{x,\theta}|\delta(x,y)=\varphi\}$,
$\mu_c$-one dimensional Lebesgue
 measure on a circle and $p_U$ ---   Lebesgue   measure
$\mu$   on the sphere normalized with respect to some
$U\subseteq \S,\ \mu(U)\ne0$.
\bigskip
       {\it Theorem 2.\/} $\bar p(S_\pm)=1.$
	  \bigskip
{\it Proof\/}.
  Let $X$ be measurable, $S_+\subset X\subseteq \S.$ Then
$$
\mu_c(X\cap c_{x,\varphi})\ge \mu_c(S_+\cap c_{x,\varphi})=
2\pi \sin\varphi \cos^2(\varphi/2).
$$
  $X_{x,\theta}=X\cap K_{x,\theta}$   is  $\mu$-measurable hence
  $$
  \mu(X_{x,\theta})=\int_0^\theta d\varphi\mu_c(X\cap c_{x,\varphi})
  \ge \pi \bigl(1-\cos\theta+{1\over 4}(1-\cos2\theta)\bigr).
  $$
Therefore
$$
p_{K_{x,\theta}}(X_{x,\theta})\ge \pol\bigl(1+
{{1-\cos2\theta}\over {4(1-\cos\theta)}}\bigr):=f_\theta,
$$
       which
  implies for any $x$
  $$
\lim_{\theta\to 0}p_{K_{x,\theta}}(X_{x,\theta})=1\eqno(*)
$$
Consider now a triangulation of the sphere,  that is,    a covering
  of $\S$  with curvilinear triangles of disjoint interiors. We must
take a sufficiently non-pathological triangulation,  for   example,
the one by means of equal equilateral triangles. Now,  any of   the
triangles can be filled with the spherical sectors
$K_{x,\theta}$        in such a
  way that their interiors are disjoint and a measure of their   sum
is arbitrarily close to the measure   of   the   triangle.   This   is
possible because   in   any   neighbourhood   of   any   point   of   the
Pitowsky sphere there   exists   a   white   point.   Let
$\tau$     be   some
triangle from the triangulation and $\{\tau_n\}$
 be a   sequence   of   sets
which are sums of the sectors
$K_{x,\theta}\subset\tau $,  for some   $x$   and  $\theta$, whose
  interiors are disjoint,  and
  $p_\tau(\tau\\ \tau_n)$ is monotonically   descending
to 0 as $n$   increases. Now,  for any $\tau$
   and any $0<\delta\ll 1$,  there   exist
$y(\delta)$ and $\theta(\delta)$ such that
$$
\mu(K_\tau)-\mu(K_{y(\delta),\theta(\delta)})<\delta,
$$
where $K_\tau$   is a spherical sector inscribed   into    $\tau$
     (this   follows
again from the fact that in any neighbourhood of a black point
there exists a white one). If we take this $K_{y(\delta)
,\theta(\delta)}$              as $\tau_1$,       then
it follows, from the
 geometry of the equilateral   triangle    $\tau$,      that,
for sufficiently small  $\delta$,
    a measure  $\mu(K)$  of any sector $K\subset\tau\\ \tau_1$       is
smaller than this of  $\tau_1$. This,  on   the   other   hand,
    means   that
$1>p_{\tau_n}(X\cap \tau_n)>p_{\tau_1}(X\cap \tau_1)=f_{\theta(\delta)}$
  for any $n>1$,    because   the   smaller   the
sector around a white point is,  the whiter are its parallels   and
the bigger is a probability measure of $X$ in this sector. The last
inequality   means   that,   if   we   take     the     finer     and     finer
triangulations (i.e., the limit $(*)$), then
$p_{\tau_n}(X\cap \tau_n)\to 1$
     for   any
$n$. The fact that $p_\tau(\tau\\\tau_n)$
is monotonically descending to 0   means,
by the same argument as above,  that   $p_{\tau_n}(X\cap \tau_n)$
   is   monotonically
ascending to  1. Therefore the smallest mesurable   set   containing
$S_+$   is of probability measure 1  on the sphere, because
$\S=\bigcup \tau$     and
$\bigcup_n\tau_n\to\tau$
    by definition. This proves that $\bar p(S_+)=1$. Reversing   the
roles of the white and black points    we   can   analogously   prove
  that $\bar p(S_-)=1$,   where $S_-  =\S\\S_+$.$\square$
\medskip
       The technique we have applied in the proof could be called   a
``rastering'' of $X$;  it means that the more   precise   the   ``raster''
is,  the whiter  the sphere. On the other hand, it   follows   that
there exists also a black ``raster,'' since in any neighbourhood   of
a white point there is a black one. Theorems 1 and 2, supplemented
with Pitowsky's law of large numbers for non measurable sets   [4],
show that the model does not predict the probability  1/2     for   an
event ``$x$ is white." Such a choice is,  however,  consistent   though
still arbitrary. This   fact   means   that   the   notions   of
{\it total
measurability}     and    {\it total   probability}
     in   Gudder's   theory     of
probability   manifolds   [3]   cannot   be     identified     with     the
probability implied by the generalized law of large   numbers   for
non measurable sets (Theorem 5.7 in [4]). The   total   probability
of $S_+$   is 1/2,    while ``with probability I equal 1"   (cf.   [4])   we
obtain any number between 0 and 1.
\bigskip
\def\c#1#2{c_{#1,#2}}

\item{\bf 3.} {\bf ON PITOWSKY'S PROOF OF\hfill\break
 NONMEASURABILITY}
\bigskip
  In [1] Pitowsky advocated the following theorem (Theorem 4):
       Let $\{S_1,...,S_n,...\}$
	   be   a   random   sequence   of   the   Pitowsky
  spheres   and   let  $\x z$
        be     fixed.     Consider     a     subsequence
$\{S'_1,...,S'_k,...\}$
 of the original   sequence   of   all   those   spheres
that satisfy  $S_n(z)=+1$. Let
$$
{\textstyle
B=B(z,\{S_n\})=\{\x w |{1\over k}\sum_{m=1}^kS'_m(w)\to w\cdot z\}.}
$$
Then
\medskip
\noindent
(a) $\mu_c(B\cap c_{z,\theta})=\mu_c(c_{z,\theta})$ for $0<\theta<\pi$,

\noindent
(b) $B$ is $\mu$-nonmeasurable.
\medskip
The proof goes as follows. Consider $w\in c_{z,\theta}$,
        for some    $\theta$,      and
  take a sequence  $\{S'_1(w),...,S'_k(w),...\}$.
     This   is   a   sequence   of
numbers  $\pm 1$,
     while a probability measure of the set of the
	white points on $c_{z,\theta}$
	     is
$\cos^2(\theta/2)$.
 Therefore, with probability 1  (in   terms   of  $\mu_c$),
$$
{1\over k}\sum_{m=1}^k S'_m(w)\to \cos\theta
$$
Let us pause here for a moment. We have assumed that a probability
space   suitable   for   this   problem   is   given   by   the   parallel
containing $w$. Note however that the two points belong also   to   a
great   circle $C_{w,z}$
        which   is   half white   and     half black.     A
conditional probability
$$
   P(``w{\rm\ is\ white"} |
   ``z{\rm\ is\ white\ and\ }\delta (w,z)=\theta ")\eqno(**)
  $$
depends not only on the measure of the set of the white points on $C_{w,z}$
       but   also
on the way the white   set   is   distributed   on   $C_{w,z}$$^2$.
   In   the
construction of the sphere (Theorem 1  in [1]) we do not assume on
$C_{w,z}$   anything but the measure of the white set. Therefore,   if   we
take $C_{w,z}$    as the probability space, we shall get $(**)=1/2$
    for   all
$w$ exept $w=\pm z$. From a different point of view it is clear that   it
is impossible to cover $C_{w,z}$       with    $\mu_c$-
measurable   sets   such   that
$(**)=\cos^2(\theta/2)$,  because then the standard argument based   on   the
Bell inequality holds.

We can see here an analogy to the Bertrand paradox   from   the
classical probability calculus: The probability   depends   on   the
choice of the probability space but this choice is not implied by
the very formulation of the probabilistic problem.

Indeed,  if we take the great circle as the probability space,
then ${1\over k}\sum_{m=1}^k S'_m(w)\to0$,
 which agrees with the Pitowsky's result only for
$\theta=\pi/2$. This implies equality   of   $B$   and   $\c z{\pi/2}$
            modulo   set   of
measure 0. Hence $B$ is measurable and  $\mu(B)=0$.

It follows that, in order to get the thesis of the theorem, one
has to introduce an additional assumption which is   by   no   means
self-evident. The theorem is false as long as   the   asumption   is
not formulated explicitely. The question of its   reasonableness   is
discussed in the next section; we shall see that it   is   possible
to modify the   model   in   a   way   which   explicitly   avoids   this
difficulty.

The Pitowsky sphere is an example of a probability   manifold
therefore the ambiguity in the choice of   the   probability   space
will find its reflection in probability   interpretation   of   such
manifolds [3].
\bigskip
\item{\bf 4.} {\bf ``PROBABILITY III"}
\bigskip
Let   us   temporarily   accept   the   Pitowsky's     choice     of     the
probability spaces   for   the   sphere.   In   order   to   distinguish
between the probabilities discussed in [4] (``probabilities I   and
II"), let us term the assumption a   ``probability   III."   It   means
that, if in some problem we can calculate probabilities   by   means
of different probability spaces, then we can decide which space to
choose and then a suitable law of large numbers holds ``with
probability III equal 1." For the Bertrand paradox,  for   example,
we can decide which probability can be realized ``with probability
III   equal 1."   However,    such   a   probability   can     be     found
inconsistent with   some   auxiliary   criterion,    as   in   Jaynes'
analysis of the Bertrand paradox [5].

As far as the   models   of   spin   are   considered,   a   standard
additional criterion   of   this   type   is   provided   by   the   Bell
inequality. So let us investigate what constraints,  if   any,    the
Bell inequality imposes on probabilities on the Pitowsky sphere.

I shall begin the discussion with a   construction   of   a   new
  pathological decomposition of $\S$
           and   a   new   version   of   the
spin-1/2     model for which the Bell inequality explicitly cannot be
derived.

 For any $\x {x,\ y}$,        let us define an equivalence relation   as
follows: $x\sim y$ iff $x=- y$. Let $Y$ be any set and a mapping
$$
f:\S/\sim\ \to Y
$$
\def\R{{\cal R}}
be bijective. Let  $\R_x=\{\c x\theta\}_\theta,\
  \R=\bigcup_x\R_x$ and $\Pi: \R\to \S/\sim$
  be defined
as  $\Pi(\c x\theta)=[x]  =\{ x,- x\}$.
Let $\{\alpha \}$ be such a well ordering of the family  $\R$
     that    according
to the continuum hypothesis  a cardinality of   any
$X_\alpha=\{c_\beta\in\R |\beta\le\alpha\}$
is ${\= X}_{\!\!\alpha}\le\aleph_0$.
 Let $Y_\alpha=\bigcup_{\beta<\alpha}(c_\alpha\cap c_\beta)$.
Let us consider a mapping $F_\beta: c_\beta\to Y,\ \beta\le\alpha$
  defined as
$$
\left.\eqalign{F_\beta(x) &= f\circ\Pi(c_\beta)\ {\rm\ iff\ }
x\in c_\beta\\Y_\beta\cr
&=f\circ\Pi(c_{\beta'})\ {\rm iff\ }x\in Y_\beta\
{\rm and\ } c_{\beta'}\ {\rm is\ the\ first\ element\ of\ }\cr
& \qquad\qquad\qquad
{\textstyle X_{\beta,x}=\bigl\{c_\gamma\in X_\beta|\bigcap_\gamma c_\gamma
\cap\{ x\}=\{ x\}\bigr\}}.\cr}\right\}
$$
  Since  $\beta\le\alpha$  is arbitrary,
      a   principle   of   transfinite   induction
states that $F_\beta$   defines a function
$$ F: \bigcup_{\beta\le\alpha}c_\beta\to Y$$
 such that
$F|_{c_\beta}=F_\beta$.
  Furthermore, as $\alpha$ is arbitrary  it follows that $F$
     is   defined   on
  the whole  $\bigcup_\beta c_\beta=\S$    and $F|_{c_\beta}=F_\beta$
     for any  $\beta$.
\bigskip
       {\it Theorem 3.\/}
\bigskip
(a) $\forall_{y,y'\in Y}(y\ne y')\Rightarrow\bigl(F^{-1}(y)\cap
F^{-1}(y')=\emptyset\bigr)$.
\bigskip
(b) $\forall_{y\in Y}\bigl(a\in f^{-1}(y)\bigr)\Rightarrow
\Bigl(\forall_\theta\ \mu_c\bigl(\c a\theta\cap F^{-1}(y)\bigr)
=\mu_c(\c a\theta)\Bigr)$.
\bigskip
(c) $\bigcup_{y\in Y}F^{-1}(y)=\S.$
\eject
{\it Proof.\/}
  \bigskip
(a) Let $y\ne y'$ and $F^{-1}(y)\cap
F^{-1}(y')\ne\emptyset$. $f$   is   bijective   hence
$f^{-1}(y)=[x]\ne f^{-1}(y')=[x']$. $\Pi^{-1}([x])=\R_x,\
\Pi^{-1}([x'])=\R_{x'}$.
   Let $z\in F^{-1}(y)\cap F^{-1}(y')$.
$\exists !_{(\varphi,\varphi')}z\in \c x\varphi\cap\c {x'}{\varphi'}$
($\exists !$  means ``exists   only   one").   $\c x\varphi$
        and $\c {x'}{\varphi'}$
    belong to $X_{\alpha,z}$       where  $\alpha$
     is an ordinal number of the greater
of   them.   Let   $c_{\alpha_0}$ be     its     first     element.     Then
$f\circ \Pi(c_{\alpha_0})=f\circ \Pi(c_{x,\varphi})=
f\circ \Pi(c_{x',\varphi'})=y=y'$. Contradiction.$\square$
\bigskip
(b) $\forall_{a,\theta}\exists_\alpha c_\alpha=c_{a,\theta}$. Moreover,
$c_\alpha\\\bigl(c_\alpha\cap F^{-1}(y)\bigr)=Y_\alpha$
and $Y_\alpha$   is
  at most countable. Therefore
  $\c a\theta\\\bigl(\c a\theta\cap F^{-1}(y)\bigr)$ is measurable
and  its measure is 0. It follows that
  $\mu_c\bigl(\c a\theta\cap F^{-1}(y)\bigr)=\mu_c(\c a\theta)$.
\bigskip
(c) It is obvious.$\square$
\bigskip
  A discussion presented below  of   the   relation   of   the   Pitowsky
model to the Bell inequality was inspired by the paper   by   Jozsa
[6] and by an earlier obsevation that the Bell inequality can   be
violated by   local   theories   admitting   random   variables   whose
domains   are   not   identical   [7].   This    further    reduces   the
Pitowsky model (at least its modified version)   to a class of
local hidden variables models with undetected signals. The   first
construction of this type goes back to a rather forgotten paper
by Pearle [9] and to an unpublished work   by   E.P.Wigner   (cf.   a
footnote in [9])  and can be   considered   as   a   hidden variables
form   of   Bohr's   complementarity   principle.   (Complementary   are
random variables whose domains are not   identical:   For   a   point
which belongs to a difference of the domains one random   variable
has a well defined value while for the   other   one   it   makes   no
sense to talk about its value.)

Let $\{S_n\}$ be   a   random   sequence   of   the   Pitowsky   spheres.
Consider a measurement of spin in a   direction   $a\in [a]=f^{-1}(y)$.
   We
shall obtain a sequence of results  $\pm 1$. Consider now a measurement
of spin in a direction $b\in [b]=f^{-1}(y')$ and assume that for any
$S_k$   a
domain of a random variable $b_k=S_k(b)$ is given   by   a   probability
sub-manifold $F^{-1}\bigl(f([a])\bigr)\cap S_k$.
 This manifold is ``almost   the   whole"
Pitowsky sphere and hence possesses the same probability   properties.
For   a   sequence   of     measurements     in     different     directions
$\{a,b,c,...\}$, that   is,  those   with   different     classes
$\{[a],[b],[c],...\}$
      the     domains     of     the     random     variables
$\{b_k,c_k,...\}$
 are disjoint (``type B random   variables"   [7]),   and
the Bell inequality cannot be   derived.   The   assumption   that   a
measurement of spin rotates a sphere guarantees a conditioning by
only   the   previous   measurement   [6].   Then,   according   to     the
``probability III" assumption, one   obtains   a   model   of   spin-1/2,
although no contradiction with the Bell Theorem is found.
\eject
\item{\bf 5.} {\bf THE AERTS MODEL}
\def\c#1#2{c_{#1,#2}}
  \bigskip
Aerts [2] has given an example of a macroscopic  classical system
in which the conditional probability $\cos^2(\theta/2)$
 occurs so we   seem
to get into conflict with the   Bell   Theorem.   The   situation   is
rather astonishing as the Bell inequality is,  in this context,    a
criterion for the Kolmogorovity   of   the   probabilistic   problem.
It   follows   therefore   that   there     exist     fairly     `ordinary'
situations where the   classical   probability   calculus   does   not
apply.

It is obvious [2,10] that there exist probability models that
are neither Kolmogorovian   nor   quantum   (Hilbertian).   A   simple
example of such a situation   was   given   by   Aerts   in   [2a].   It
would be therefore interesting to find out which elements of   the
Aerts model are    {\it sufficient\/} for   the   non-Kolmogorovity   of   the
description. We shall see that a    {\it conditioning   by   a   change   of
state\/} is,    actually,    the   required   sufficient   condition.     No
form of the Bell inequality can be derived   for   such   models   as
should be clear from   the   hidden variables   description   of   the
Aerts model given in Sec. 6. There exists,  however,  also   much
simpler argument for the non-Kolmogorovity   of   such   models.   It
will be shown,  in the fifth variation on the   Aerts   model,    that
the acts of conditioning by a change of a state in general  {\it
do not
commute}.

The question whether   one   is   capable   of   constructing   the
{\it quantum\/}   probability model based on the conditioning by   a   change
of state as the unique non classical element shall be   left   open
in this paper. As long as the concrete Aerts model   is   concerned
 {\it two\/}   assumptions are needed: the conditioning   by   a   polarization
and a lack of knowledge about a measurement. It is still   unclear
for the author of this paper   whether   the   latter   condition   is
sufficient for the non Kolmogorovity of the description.

We consider a particle with a   positive   charge   $q$   which   is
located on a sphere. The measurement consists   of   the   following
operation: We take two negative   charges   $q_1$     and   $q_2$
     such   that
$q_1+q_2  =Q$ and locate them on antipodal points on the sphere; $q_1$
     is
chosen at random between 0 and Q and we give the   outcome +1  to
the measurement if the Coulomb force  $|F_1|$   acting between $q_1$
   and $q$
is greater that this,  $|F_2|$,    between $q_2$   and $q$.
   Otherwise   we   give
the outcome  $-1$. We find that the probability $P_+(\theta)$
 of the outcome
$+1$   is equal to $\cos^2(\theta/2)$
where    $\theta$ is the angle between $q_1$     and   $q$.
We assume also that during the measurement the charge $q$ moves   in
a direction of the Coulomb force and,  after   the   measurement   is
completed,  remains at the location point of either $q_1$   or $q_2$
   (this
depends on the result of the measurement); we obtain in this   way
a change of a state of the system analogous   to   this   due   to   a
Stern-Gerlach device or an optical polarizer.   A   reader   who   is
afraid that the charges can annihilate,  may think of   masses   and
the Newton force.

>From   a   hidden variable   point   of   view   a   result   of   the
measurement is fully determined by a pair of hidden   variables
$(q_1,\alpha)$   where  $\alpha$
   denotes an ``angle of polarization" of the charge
$q$ with respect to some   experimental   device   consisting   of   the
charges $q_1$   and $q_2$. $P_+(\theta)$
 is the   probability   of   the   result +1
provided the hidden variable   $\alpha=\theta$
    or,    because   of   the   assumed
polarization mechanism,  provided a previous experiment ---  with the
measuring device suitably directed ---  has   given   an   outcome  +1.
Note,  however,  that, if we make another measurement with the
{\it same}
  device sloped by  $\alpha\le\theta$,   then
with probability 1  we shall obtain the   same   result   as   before.
Therefore we must assume   that   either   the   pair   $(q_1,q_2)$
    gets
distributed at random after the measurement   or   we   simply   take
another device. Both conditions are referred to as   the    {\it
lack   of
knowledge about the measurement}.

 The pair of charges $(q_1,q_2)$ put diametrically on   the   sphere
plays a role of   the   measuring   device;   we   shall   denote   this
measurement by  $e_{\alpha,\beta}$ with $(\alpha,\beta)$
 being polar coordinates of $q$.   We
assume that before the first   measurement   $q$   is   distributed   at
random on the sphere with equal probability in   every   direction.
In   the   discussion   we   will   simplify   our   considerations     by
constraining all the charges to one great circle on   the   sphere.
Therefore instead of the pair $(\alpha,\beta)$ we can write $(\alpha,0)$.

For any measurement  $f=e_{\alpha,0}$        we write
	  $f=f_1$     if the result   of  $f$
is +1   and  $f=f_2$
     in the opposite case. We find $P(f=f_1)=P(f=f_2)=1/2$.

In   the   following   variations   we   shall   be   dropping   some
assumptions about the model. In this way we   shall   localize   the
element which is sufficient for the   non-Kolmogorovity.
  \bigskip
{\bf (a) First Variation.\/}
         Let us consider the Aerts' model but assume that   during   and
after the   measurement   the   charge   $q$   remains   in   its   initial
position. In   other   words,    we   drop   the   assumption   that   the
measurement polarizes the measured system.

Now let $P(f=f_1|g=g_1)$ denote the probability that  $f=f_1$    if the
state of the system is such that an eventual measurement  $g$   (where
$g$   represents the  {\it same}
   device but rotated   by   some   angle)   would
yield  $g_1$.

The probabilistic nature of this situation arises because   we
do not know the state  of the    {\it whole     system}   (=   measuring
device +  charge $q$). We shall get  $g=g_1$     if
$$
q_1(g)>Q\sin^2\bigl(\theta(g,q)/2\bigr),\eqno(6)
  $$
where we have denoted by $q_1(g)$
 a hidden variable of the device $g$
and by   $\theta(g,q)=\theta_g-\theta$
  a polar angle between the points   of   location
of $q_1$   and $q$. Analogously we shall obtain  $f=f_1$    if
$$
q_1(f)>Q\sin^2\bigl(\theta(f,q)/2\bigr).\eqno(7)
$$
Equations (6) and (7) imply the following conditional probability
$$
P(f=f_1|g=g_1)=1-{2\over\pi} {\rm sign}(\theta_g-\theta_f)\sin
\bigl((\theta_g-\theta_f)/2\bigr),\eqno(8)
$$
which is Bayesian (Kolmogorovian).
\bigskip
{\bf (b) Second Variation.}
 Consider the same situation as in   the   first   variation   but
assume that the measurements  $f$   and  $g$
   do   not   correspond   to   the
{\it same}   rotated experimental device.   In   the   first   variation   the
first measurement imposes a limitation on possible   localizations
of the pair $(q_1,\alpha)$ in $[0,Q]\times[0,2\pi)$.
   If   we   take   two    {\it different\/}
devices corresponding to  $f$ and  $g$,   then a measurement made by means
of  $g$
     imposes   no   limitations   on   the   possible   outcome   of   a
subsequent measurement made by means of  $f$. This follows from   the
fact that there is a lack of knowledge about the   hidden-variable
states (given   by   $q_1$)   of   the   two   devices   and   their   hidden
variables are not related to each   other.   Therefore   the   second
measurement is independent of the first one, and we get
$$
P(f=f_1|g=g_1)=1/2.\eqno(8')
$$
This probability is also Bayesian.
\bigskip
{\bf (c) Third Variation.}
  Let us assume that there is no   effect   of   polarization   but
that we know a location of $q$. Then
$$
P(f=f_1)=\cos^2(\alpha/2),
$$
where  $\alpha$   is the polar angle between $q$   and   $q_1$.
   This   probability
{\it looks\/}   non-Kolmogorovian.
 However  this   is   not   the    {\it conditional\/}
  probability as the   ``condition''
     that   appears   here   is:   ``$q$   is
located at a given point." A probability of the   condition   is   0
and the Bayes rule does not apply trivially because   we   are   not
allowed to   divide   by   0   (cf.   the   Borel   paradox   [8]).   This
probability satisfies all the   postulates   of   the   Kolmogorovian
calculus.
\bigskip
{\bf (d) Fourth Variation.}
  Consider the same situation as   in   the   previous   point   but
assume that the conditioning is by ``$q$ belongs to some   given   set
of a nonvanishing measure." This set   can   be   taken   arbitrarily
small  so that the model should lead to probabilities arbitrarily
close to these from the previous point. One may expect   that   the
model is non-Kolmogorovian as   the   Borel-like   objection   is   no
longer valid.

We can try to prove its supposed non-Kolmogorovity in   a   way
analogous to this from the Aerts paper [2a].

Let us assume that the charge $q$ is located   in   an   interval
that is  $\theta\in[\alpha-x,\alpha+x]=I_{\alpha x}$,
     instead of the location in a point,   i.e., $\theta=\alpha$.
 We find that
$$
P(f=f_1|\theta\in I_{\alpha x})=\pol \bigl(1+\cos(\theta_f -\alpha)
{\sin x\over x}\bigr).\eqno(9)
$$
For $x=0$ we get the required $\cos^2\bigl( (\theta_f-\alpha)/2
  \bigr)$; for   $x=\pi$     we   get
$P(f=f_1)=1/2$. This model can be   made   arbitrarily   close   to   the
third variation and no conditioning by an event whose probability
is 0 appears. The contradictory constraints derived by Aerts are,
however,  no longer valid. Let us follow the Aerts' reasoning.

Assume that there exists a Kolmogorovian model. We have
$$
\eqalign{\mu (F_1\cap G_1) &=P(f=f_1|g=g_1)P(g=g_1)\cr
&=
P(f=f_1|\theta\in I_{\theta_gx})P(\theta\in I_{\theta_gx})\cr
&=\pol \bigl(1+\cos(\theta_f -\theta_g)
{\sin x\over x}\bigr){x\over \pi}\cr
&=\mu(E_1\cap F_1\cap G_1)+\mu(E_2\cap F_1\cap
 G_1),\cr}\eqno(10)
 $$
 $$
\eqalign{ \mu(E_1\cap G_1)&=\pol \bigl(1+\cos(\theta_e -\theta_g)
{\sin x\over x}\bigr){x\over \pi}\cr
&=\mu(E_1\cap F_1\cap G_1)+\mu(E_1\cap F_2\cap
 G_1).\cr}\eqno(11)
  $$
Following Aerts, we take  $\theta_f-\theta_g=\pi/3$,
$\theta_e-\theta_g=2\pi/3$, then
$$
\mu(F_1\cap G_1)=\pol (1+\pol {\sin x\over x}) {x\over \pi},
$$

$$
  \mu(E_1\cap G_1)=\pol (1-\pol {\sin x\over x}) {x\over \pi},
  $$
and $\mu(E_2\cap F_1\cap G_1)\ge \sin x/2\pi$. According to $(8')$, we have
$$
\mu(E_2\cap F_1)={1\over 4}.
$$
Hence $\mu(E_1\cap F_1\cap G_1)\le 1/4$.
 These constraints are   not   contradictory.
The point is that   we   practically   deal   here   with   independent
events if we assume that different measurements are made by means
of different (i.e., independent) devices   (cf.   Eq.$(8')$).   On   the
other hand,  if we assume that we make measurements   by   means   of
the same   device,   then   we   will   deal   only   with   Kolmogorovian
probabilities  like those given by (8).
\bigskip
{\bf (e) Fifth Variation.}
 Let the charge $q$ be distributed at random on   a   great   circle
and let the measurement polarize the system. Assume,  in addition
that both $q_1$   and $q_2$   are known. For $q_2\ne 0$,
the probability   of    $f=f_1$
in the first measurement is
$$
P(f=f_1)={1\over\pi}{\rm arctg}\bigl(\pol\sqrt{q_1/q_2}\bigr).
$$
Let now the charges satisfy
$$
{\pi\over 2}<{\rm arctg}\bigl(\pol\sqrt{q_1/q_2}\bigr)<\pi.
  $$
Assuming that  $f=e_{0,0},\ g=e_{\pi/2,0}$,     and  $h=e_{\pi,0}$
        we find
$$
  P({\rm first\ }f=f_1\ {\rm then\ }g=g_1\ {\rm and\ finally\ }
  h=h_1)=P(f=f_1),
$$
while
$$
P({\rm first\ }f=f_1{\rm \ then\ }h=h_1)=0.
$$
The   probability   depends   here   on   the   order   of   measurements,
and,  of course,  no model based on a probability space can lead to
such a result.

The   last   variation   shows   clearly   that   in   models     with
 {\it conditioning by a change of state\/}
      a   probability   of    {\it subsequent\/}
  events depends in general   on   the   order   of   measurements.   The
calculus is therefore in general non-Kolmogorovian (and non-Boolean).
 An exeptional
situation   arises   only   in   case   the   events   in   question   are
independent.
\bigskip
\item{\bf 6.} {\bf A HIDDEN VARIABLES DESCRIPTION\hfill\break
OF THE AERTS MODEL}
\bigskip
  A hidden variables description of   the   model   is   the   following.
(Again,  for simplicity,  I assume that the charges $q$,      $q_1$
     and   $q_2$
are constrained to one great circle on the sphere ---  it in no   way
restricts the generality of the problem.)

Before the first measurement the whole system is described by
a probability density
$$
\rho_0(q_1,\phi)={1\over 2\pi Q},
$$
defined   on   a   probability   space   of     the     hidden     variables
$\Lambda=[0,Q]\times[0,2\pi)$.

A random variable representing spin measurement by a device $A_\alpha$
  (i.e., the one with $q_1$   whose polar coordinate is  $\alpha$)
   is defined as
$$
\eqalign{A_\alpha(q_1,\phi) &= +1\ {\rm iff}\ q_1>Q\sin^2\bigl(
 (\alpha-\phi)/2\bigr)\cr
 &=-1  \ {\rm iff}\ q_1\le Q\sin^2\bigl(
 (\alpha-\phi)/2\bigr).\cr}
 $$
A probability of the result +1   in the first measurement is
$$
\eqalign{
P(A_\alpha=+1)&=\int_\Lambda d\lambda\rho_0(\lambda)
\chi_{\alpha,+}(\lambda)\cr
&={1\over 2\pi Q}
\int_0^{2\pi}d\phi\int_{Q\sin^2\bigl(
 (\alpha-\phi)/2\bigr)}^Qdq_1=\pol,\cr}
$$
where  $\chi_{\alpha,+}$
        is a characteristic   function   of   the   set   of   those
hidden variables for which $A_\alpha(q_1,\phi)=+1$.

       The first measurement changes the probability distribution of
the hidden variables:
$$
{\vcenter{\halign{$#$&$#$&$#$\hfil &$#$\hfil\cr
\rho_0 &\to \rho_{\alpha,+}(q_1,\phi)&=Q^{-1}\delta (\alpha -
\phi),\ &{\rm if\ the\ result\ was\ }+1;\cr
&\to \rho_{\alpha,-}(q_1,\phi)&=Q^{-1}\delta (\alpha -
\phi+\pi),\ &{\rm if\ the\ result\ was\ }-1.\cr}}}\eqno(12)
$$
Hence
$$
\eqalign{
P(A_\beta=+1|A_\alpha=+1)&=\int_\Lambda d\lambda\rho_{\alpha,+}
(\lambda)
\chi_{\beta,+}(\lambda)\cr
&={1\over Q}
\int_0^{2\pi}d\phi\ \delta(\alpha-\phi)\int_{Q\sin^2\bigl(
 (\beta-\phi)/2\bigr)}^Qdq_1\cr
 &=\cos^2\bigl( (\alpha -\beta)/2\bigr).\cr}
 $$
Formulae (12) provide a hidden variables description of   a ``{\it reduction
of a wave packet\/},"   which physically means that the charge   $q$   falls
down on either $q_1$   or $q_2$.
 The measurement polarizes the system and
by means of the polarization process, fixes the hidden variable:
$\phi\to\alpha$.

Is the system of charges on the sphere a Kolmogorovian one or
not?

It depends. If we ask ``What is a probability that the   system
is in such a state that  {\it eventual\/}   measurements of
$f$     and    $g$     would
yield  $f=f_1$     and  $g=g_1$?"
then we shall get the result   as   in   Sec.5
(Eq.(8)   or   $(8')$),    and   the   system   is     Kolmogorovian.     The
conditional probability obtained in this   way   does   satisfy   the
Bell inequality although this is not this conditional probability
that can be tested experimentally.

A completely different situation arises if we ask   ``What   is   a
probability of getting  $f=f_1$     provided in an earlier measurement we
have obtained $g=g_1$?" This is the problem   one   faces   in   actual
experiments. The conditional   probability   derived   in   this   way
yields   the   correct   (and   non-Kolmogorovian)   spin     model     if
subsequent measurements are done by means of independent devices.
There is no contradiction with the Bell Theorem,    because   it   is
impossible to derive the Bell   inequality   for   this   model.   The
polarization       mechanism       makes  {\it alternative}
         measurements
{\it complementary\/}.
     We   therefore   obtain   another     hidden variables
representation of the complementarity principle $^2$. A   difference
between the two questions is   exactly   this   between   the   models
without or with the polarization.

The       fifth   variation and the hidden variables   description
above suggests that the very fact of  {\it conditioning by   the   change
of state\/}   is non-Kolmogorovian. Indeed,  in Bayesian conditioning a
number of particles with a given property in a sample is the same
both before and after the act of conditioning   by   this   property
(``{\it conditioning by filtering\/}''). In the Aerts' example   the   sample
may not contain the charges $q$ located at some given point   before
the measurement (probability of such a localisation   is   0),    but
after the first measurement about half of them is at this   point.
``A reduction of a wave packet" in the   Kolmogorovian   probability
takes   always   a   form
$$
{\textstyle
\rho \to \rho_{\alpha\pm}=\rho/\int_\Lambda\rho(\lambda)
\chi_{\alpha,\pm}(\lambda)d\lambda},
$$
where   the
  characteristic function corresponds to a   set   whose   measure   is
{\it non zero}.
   In   polarisation   phenomenon   (understood     at     least
clasically, i.e., in terms of the   hidden   variables)   the   act   of
measurement turns  with a non zero probability  the state of   the
system into a state whose probability before the measurement   was
zero. And this is clearly a non-Kolmogorovian behaviour.
\bigskip
\noindent{\bf ACKNOWLEDGEMENT}
\bigskip
  I would like to thank Zbigniew   Mielewczyk   for   his   stimulating
remarks  and the referee for calling my attention to Ref.[2b].
\bigskip
\noindent{\bf REFERENCES}
\bigskip
\item{1.} I. Pitowsky,  {\it Phys. Rev. D\/} {\bf 27},   2316 (1983).\par
{\leftskip=-2.8pt
\itemitem{2. (a)} D. Aerts, {\it J. Math. Phys.\/} {\bf 27}, 202 (1986).
\itemitem{\ \ \ (b)} D. Aerts,  ``The origin of the non classical character of
 the
quantum probability   model,"    in   {\it Information,    Complexity    and
Control in Quantum Physics\/},   A. Blanquiere,  S.   Dinier,
and G. Lochak, eds.  (Springer,  New York, 1987).
\itemitem{\ \ \ (c)} D. Aerts,  {\it Helv. Phys. Acta\/} {\bf 64}, 1 (1991).\par}
\item{3.} S. P. Gudder,  {\it J. Math. Phys.\/} {\bf 25}, 2397   (1984).
\item{4.} I. Pitowsky,  {\it Quantum Probability --- Quantum   Logic\/},
    Lecture
Notes in Physics {\bf 321}    (Springer, New York, 1989).
\item{5.} E. T. Jaynes,  {\it Found. Phys.\/} {\bf 3}, 477 (1973).
\item{6.} R. Jozsa,  {\it Found. Phys.\/} {\bf 19}, 1327 (1989).
\item{7.} M. Czachor,  {\it Phys. Lett. A\/} {\bf 129}, 291  (1988).
\item{8.} P. Billingsley,  {\it Probability and Measure\/}  (Wiley,
   New York, 1979)
\item{9.} P. M. Pearle,  {\it Phys. Rev. D\/} {\bf  2}, 1418 (1970)
\item{10.} B. Mielnik,  {\it Commun. Math. Phys.\/} {\bf 9}, 55 (1968)
 \bigskip
\noindent{\bf NOTES}
\bigskip
\noindent
1. Permanent   address:   Laboratory   of   Dielectrics   and   Organic
Semiconductors,  Technical   University   of   Gda\'nsk,
Majako\-wskie\-go 11/12,        Gda\'nsk,  Poland.
\medskip\noindent
 2.   For   example,    let
$\forall_{C_{w,z}}\exists_{x\in C_{w,z}}\mu_c\bigl(\{ y\in C_{w,z}\cap S_+|
\delta(x,y)\le\pi/2\}\bigr)$$=\mu_c(C_{w,z}\cap S_+)$$=\pol
\mu_c(C_{w,z}).$
   Pitowsky   spheres   with   white     points
distributed on $C_{w,z}$  in this way   exist.   The   proof   is   exactly
analogous to this of Theorem 1  in [1].   Then   $(**)=1-\theta/\pi$,
    if   one
takes $C_{w,z}$
       as the probability   space.   This   example   seems   very
instructive.
\medskip  \noindent
3. If the charge $q$ falls down on some point then it   clearly   has
not fallen down on another   one.   Having   given   a   result   of   a
measurement we cannot,   within the model with   polarization,    talk
in a sensible way about its   alternative:   We   can   think   either
about  {\it successive\/}
   measurements (then the Bell   inequality   is   not
derivable) or ask ``What would have happened if...,"   but   then   we
deal with a different problem (in Aerts' terminology [2b] this is
an    {\it observation\/})
   and   obtain   again   Eq.   (8)     (there     is     no
complementarity  but there is no model of spin either).

\bye